\pdfoutput=1
\documentclass[12pt,a4paper]{article}
\usepackage{lmodern}
\usepackage[T1]{fontenc}
\usepackage[utf8]{inputenc}
\usepackage[margin=2.8cm]{geometry}
\usepackage{amsmath,amssymb}
\usepackage{graphicx}
\usepackage{booktabs}
\usepackage{microtype}
\usepackage[hidelinks]{hyperref}
\usepackage{enumitem}
\usepackage{setspace}
\usepackage[round]{natbib}

\graphicspath{{figures/}}

\title{\textbf{Reading the Count:\\ Prediction Markets and Brazil's 2026 Presidential Election}}
\author{Raphael Coelho\\ \normalsize São Paulo School of Economics, FGV}
\date{October 2026}

\begin{document}
\maketitle

\begin{abstract}
\noindent
We rebuild the public count of Brazil's 2026 presidential runoff minute by minute from the
electoral court's section-level records and compare it with every trade on Kalshi and
Polymarket. A projection published live on the night gave the eventual winner, Lula, a
probability of 0.998 from 19:15, with 1.6\% of sections counted. The markets reached 0.9 at
20:24, with 46\% counted. Read as the projection's central estimate with a wider uncertainty,
their prices imply a standard deviation of the final margin about four times the live
projection's, and about twice that of a section-level projection calibrated on the 2018 runoff,
the earlier night with the largest errors. Valued at these projections, takers' purchases of
Lula after the count resumed had an expected profit of 4 to 18 cents per Kalshi contract after
fees; valued at a state-level extrapolation of the count, it was zero. During a two-hour
suspension of the count the markets gave Lula 0.25 to 0.44, below every reading we compute
afterwards from the sections then public, none of which was available live. On the first-round
night, when the headline count showed the eventual
leader ahead from its first file, the same markets were faster than our projection.
\end{abstract}

\section{Introduction}\label{sec:intro}

After the polls close, an election market prices an outcome that is already fixed, from a
count released in an order correlated with the result. How quickly the price reaches the
outcome is a measure of how quickly the market processes the public count. On the night of
Brazil's 2026 runoff, Kalshi and Polymarket processed it slowly. Their prices implied about four
times the uncertainty of a projection published live, and buying the eventual winner at those
prices had a positive expected return under each projection we use, though not under a simple
state-level extrapolation of the count.

Brazil allows an exact version of this measurement. For the 2026 runoff the electoral court
(TSE) published, for each of 499,205 polling sections, the vote, the minute its tally sheet was
printed and the second the court received it. The national result file with $n$ sections is
reproduced by the first $n$ sections in order of receipt, so we know, minute by minute, which
sections were public. We observe every trade on both venues.

We use two benchmarks. The first is a projection that a Brazilian website, Arvor, published live
from the TSE's municipal result files every five minutes. It shows what a reader of the count
could know in real time. The second is our own projection, which compares each public section
with its own first-round vote. It uses section records published after the night, so it
measures what the public count contained rather than what a trader could compute. It also
covers the suspension of the count, before the live projection began. We calibrate its
uncertainty on the two earlier runoffs with section records, 2018 and 2022.

The live projection gave Lula 0.998 at 19:15 with 1.6\% of sections counted. The markets passed
0.9 at 20:24 with 46\% counted. Against the live projection, their prices imply a standard
deviation of the final margin 3.7 to 3.9 times the projection's own. Against the section-level
projection calibrated on 2018, the ratio is 2.1. Takers bought Lula at an average of 0.78 from
19:06 to 20:24. Valued at the live projection's probability, purchases from 19:15 had an expected
profit after Kalshi's fee of 18 cents per contract, and 4 cents with the projection's band
tripled. Valued at a state-level extrapolation, whose timing the markets matched, it was zero.

Two other results qualify this. During the suspension of the count from 17:07 to 19:00, the
markets priced Lula at 0.25 to 0.44. The sections then public, read afterwards from the section
records, gave him 0.57 to 0.92 across calibrations. On the first-round night the headline showed the eventual leader ahead from its
first file, and the markets passed 0.9 an hour before the section-level projection, whose
calibration there rests on a poorly matched pair of earlier elections.

\paragraph{Related literature.} Prediction markets aggregate campaign information well
\citep{berg2008, wolfers2004} and are well calibrated close to expiry but compressed toward one
half at long horizons \citep{page2013}.
Across a million forecasts, \citet{dickerson2026} find prediction platforms well calibrated
overall, with systematic errors in some domains, and \citet{le2026} finds political prices on
Kalshi and Polymarket compressed toward one half. Those
studies compare prices with outcomes; we compare them with the information public at each
minute. \citet{auld2019} study betting and currency markets on the night of the Brexit
referendum, \citet{snowberg2007} use election-night prices to measure partisan effects, and
\citet{croxson2014} measure how fast betting prices absorb goals. Prices that move in the right
direction but too little on strong evidence resemble the underinference from strong signals
documented by \citet{augenblick2025}; a single night cannot test their account.

\section{Setting and data}\label{sec:data}

\paragraph{The election.} Lula (PT) and Flávio Bolsonaro (PL) contested the runoff on 25
October 2026. In the first round, on 4 October, Flávio finished first by 2.0 points of
two-party margin against polls that had Lula ahead. In the runoff the final polls had Flávio
ahead by 1.6 points and Lula won by 3.57 points (51.79\% to 48.21\%), the largest poll miss of
the seven runoffs since 2002. All margins below are Lula minus Flávio, in points of the
two-party vote, and all times are Brasília time (UTC$-3$).

\paragraph{The count.} When voting ends, each section's machine prints a tally sheet
(\emph{boletim de urna}, BU). Copies are posted at the section and given to party poll
watchers, and the machine's media is carried to a transmission point. On the night the TSE
serves national, state and municipal result files, each version stamped with its generation
time; an archive kept every version on both 2026 nights. Section-level records, with printing
and receipt times, are published afterwards. We call the sections in the latest national file
the \emph{public set} and Lula's two-party share in that file the \emph{headline}. In the runoff
the version with $n$ sections appears a median of 22 seconds after the recorded receipt of the
$n$-th section (middle half: 53 seconds before to 32 seconds after), and its totals equal those of the first $n$ sections by receipt. In the first round
the national file stopped updating from 18:49 to 20:05 while the state files continued, so for
that night the public set is the sum of the state files when it is larger.

\paragraph{Markets.} Kalshi and Polymarket listed winner contracts for the runoff and for first
place in the first round. We use every Kalshi trade (92,182, with the taker's side), every
Polymarket fill with both wallets (379,943), Polymarket order-book snapshots and one-minute
prices. The price $p_t$ is the bid-ask midpoint of the Lula contract at the end of minute $t$, from
Kalshi's quotes and from Polymarket's one-minute price history; in 68 order-book snapshots on
runoff night, Polymarket's history matches the book's midpoint (median difference 0.03 cents). Kalshi charges takers $0.07\,p(1-p)$ per
contract; we assume no Polymarket fee. A purchase of Polymarket's Flávio contract is
economically a sale of Lula, so we measure Polymarket flows in \emph{Lula-equivalent shares}.
Between 16:30 and 22:30 on runoff night the venues traded 19.1 million Kalshi contracts and 46.4
million Polymarket shares.

\paragraph{Access from Brazil.} On 23 April 2026 the National Monetary Council barred contracts
on non-economic events. On 24 April the government ordered 27 platforms, Kalshi and Polymarket
among them, blocked, the Finance Ministry calling them illegal betting, and the regulator
Anatel sent the domains to internet providers.\footnote{Bloomberg Línea, ``Bloqueio a sites de
mercados preditivos coloca em xeque planos da Kalshi para o Brasil'', 27 April 2026; Gazeta do
Povo, ``Por que sites de previsões como o Polymarket foram bloqueados no Brasil?'', 27 April
2026; Capital Aberto, ``Tentativa de bilionária de trazer a Kalshi ao Brasil é bloqueada'', 27
April 2026.} Access through private networks was reported to continue.\footnote{BNL Data, 11 September 2026,
\url{https://bnldata.com.br/brasil-e-2o-em-temas-apos-bloqueio-do-governo-federal-e-segue-acessivel/}.} We cannot observe where traders were.

\section{Before the vote}\label{sec:polls}

We aggregate 193 national runoff polls with a local-level Kalman filter on the margin, with
house effects shrunk toward zero. From 8 to 24 October the markets gave Lula about 0.18. The
aggregator, with a systematic poll error of 3 points, gave 0.29. Adding the average final-week
runoff poll error from 2002 to 2022, which underestimated the right by about one point, gives
0.16; using only 2018 and 2022 gives 0.39; predicting the runoff error from the 2026 first-round
error gives 0.19. The market sat near the bottom of this range.

\section{Runoff night}\label{sec:night}

\subsection{Benchmarks}

\paragraph{The live projection.} Arvor, a site that described itself as favourable to Flávio,
committed its projection code at 16:04 on election day. From 19:15, once 1\% of sections was
counted, it published a projection every five minutes. For each municipality it compares the
runoff vote with the first-round vote at the same stage of counting, and it reports a projected
national share with a 90\% band. We convert the band to a standard deviation and the share to a
probability; its projected margin plays the role of $M_t$ below. The band contained the final
result in all 53 snapshots of the night. With a single count, this is a check on this night
only.

\paragraph{The section-level projection.} At minute $t$, for each municipality, we compute the
swing as Lula's two-party share in its public sections minus the share those sections gave him
against Flávio in the first round. We shrink the municipal swing toward the state swing (weight
$n/(n+2{,}000)$ on the municipality, $n$ its public votes) and the state swing toward the
national one (weight $n/(n+20{,}000)$), and apply it to the first-round shares and turnout of the
sections not yet public. Its projected margin is $M_t$.

For its uncertainty, each state-minute of a reference runoff yields one projection error,
indexed by the share $x$ of the state's sections then counted. The root mean square error by
$x$, made non-increasing in $x$ and bounded below by a minimum state error of 0.0005 to 0.002
of two-party share, is $e(x)$; the minimum keeps the band from collapsing late in the night. The
national standard deviation of the margin is $200\sum_s w_s e(x_s)$, where $w_s$ is state $s$'s
share of the national vote and 200
converts a two-party share into points of margin; summing treats state errors as perfectly
correlated.

\paragraph{Reference runoffs.} Calibrated on the 2022 runoff, the band was too narrow in 2026:
early in the night the 2026 state errors were about twice those of 2022, and the 90\% band
excluded the final margin in 43\% of minutes between 17:07 and 21:30. We report this calibration
with its standard deviation multiplied by $k=1,3,5$. The only other runoff with section records,
Haddad against Bolsonaro in 2018, lacks receipt times, so its order of counting is unknown. We
simulate it. Within each state, a section's rank in the order of receipt is the sum of a
municipality effect, a polling-place effect and a section term. Their variances, and how far
each effect persists from one runoff to the next, come from 2022 and 2026; the persistence
correlation is 0.46 for municipalities and 0.30 for polling places. We draw 20 orders and take
the median error curve. Applied to 2022 and 2026, each simulated from the other night, the
method understates the errors of the actual order by a factor of 1.1 to 2.9 between 0.5\% and
70\% counted, so the 2018 errors are if anything larger than we report, and the 2018 ratio and
profits below are if anything overstated. Up to half counted they
are four to eight times those of 2022. In 2018 the two finalists took 75\% of the first-round
vote, against 92\% in 2022 and 2026, so the runoff vote was a weaker function of the first-round
vote. A trader unsure which night 2026 would resemble could reasonably have used the 2018 band.
The 2026 errors fell between the two, closer to 2022's.

\subsection{The suspension}\label{sec:pause}

\begin{figure}[t]
\centering
\includegraphics[width=\textwidth]{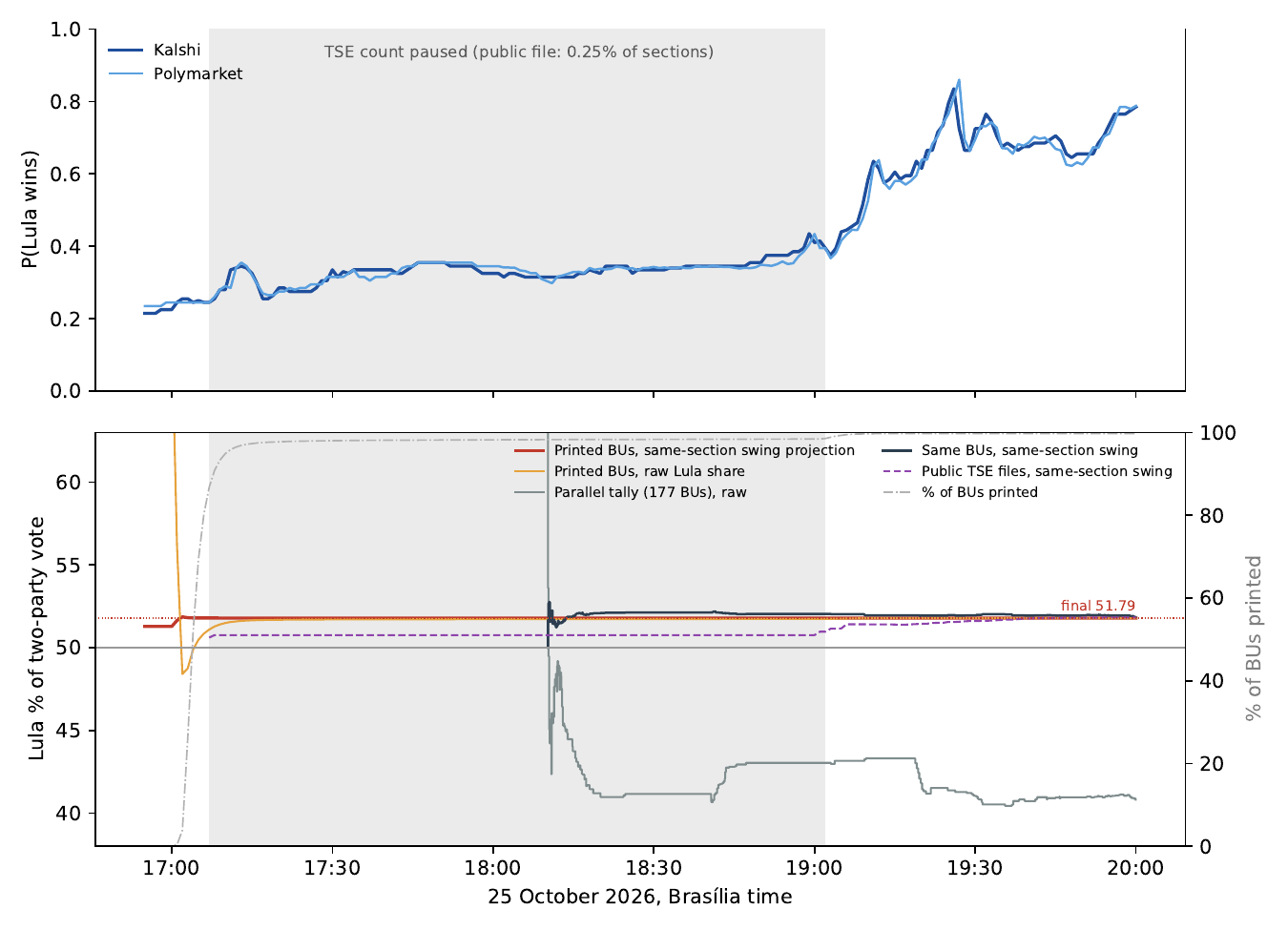}
\caption{The suspension. Top: market probabilities. Bottom: Lula's two-party share implied by
the printed tally sheets (BUs), by the volunteer tally (``parallel tally''; raw, and as a swing
against the same sheets' first-round vote) and by the public set; the share of sheets printed.
Before 18:15 the volunteer series rests on fewer than ten sheets.}\label{fig:pause}
\end{figure}

Both campaigns asked the TSE for more voting time during the afternoon, Lula's citing failures
of public transport and Flávio's citing rain. Between 17:01 and 17:03 the Supreme Court extended
voting to 19:00 where sections were still open, and the press reported it within minutes. The
TSE stopped publishing at 17:07. The national file generated at 17:06, with 1,225 sections
(0.25\%), 776 of them abroad, remained the public set until 19:00:50. Its headline showed Lula at
57.7\%, because it consisted mainly of votes cast abroad. Figure~\ref{fig:pause} shows what else
could be known.

\begin{itemize}[leftmargin=1.2em]
\item \emph{The public set.} Its sections project Lula ahead by 1.5 points, or 1.8 using only
the 449 domestic sections. That is a probability of 0.92 at $k=1$, 0.68 at $k=3$, 0.61 at $k=5$
and 0.57 under the 2018 calibration. Combined with the polls (margin $-1.6$, standard deviation
3), it is 0.87 at $k=1$ and 0.47 at $k=3$. This reading uses the section records. Live, a
trader had the municipal files, and Arvor's municipal method declined to project below 1\%
counted.
\item \emph{Printed sheets.} By 17:05, 69\% of sheets had been printed and by 17:10, 95\%. Their
section-level projection gives a margin of about 3.5 points. The sheets were posted at each
section and held by poll watchers; no one collected them centrally.
\item \emph{A volunteer tally.} From 18:10 Arvor published photographs of sheets sent in by
volunteers, 77 by 19:00 from 26 municipalities and 177 in all, each matching the TSE file. Their
raw share was 43\% Lula, but the same sections had given him 40\% in the first round. Measured
as a swing, the 77 sheets imply a Lula share of 52.0\% (clustered standard error 0.24 points, ignoring
that the municipalities were self-selected), against a final 51.8\%. Arvor reported only the
raw share; its live stream had 228 views.
\end{itemize}

From 17:07 to 19:00 the markets moved between 0.25 and 0.44, below every reading above. From
0.25 at 17:05 they rose to 0.34 by 17:30, after the extension order and the last file before
the suspension, and reached 0.44 only in the last minutes before publication resumed. Over the two hours
takers sold Lula on net, about 230,000 contracts on Kalshi and 188,000 Lula-equivalent shares on
Polymarket, so prices rose through revisions of resting quotes.

\subsection{After the count resumed}

\begin{figure}[t]
\centering
\includegraphics[width=\textwidth]{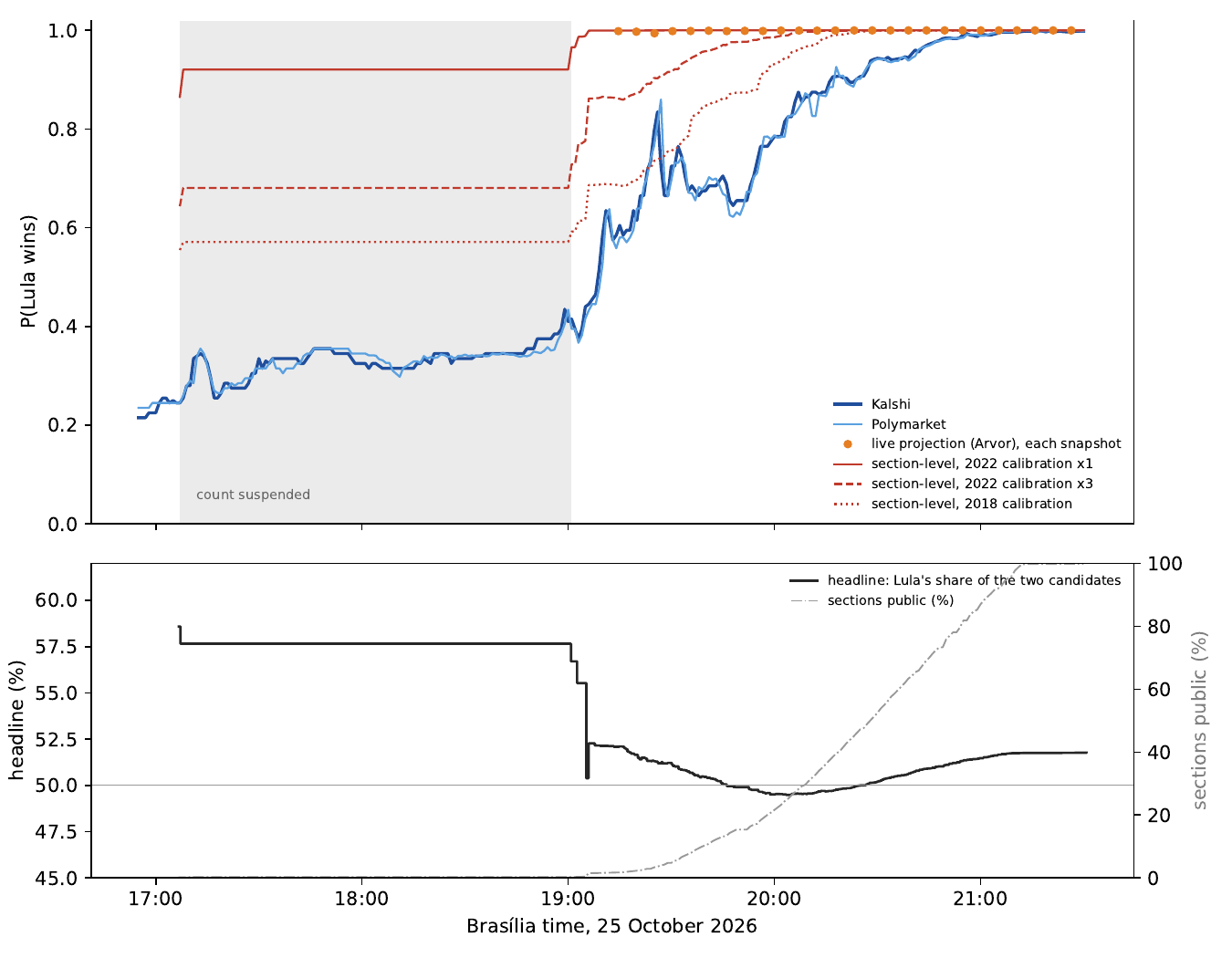}
\caption{Runoff night, 25 October 2026. Top: market probabilities, the live projection at each
snapshot, and the section-level projection under the 2022 calibration ($k=1$ and 3) and the 2018
calibration. Bottom: the headline (Lula's two-party share in the public set) and the share of
sections public. Shaded: the suspension of the count.}\label{fig:runoff}
\end{figure}

Figure~\ref{fig:runoff} and Table~\ref{tab:night} show the rest of the night. The live
projection put Lula at 52.8\% (band 51.2 to 54.3) at 19:15, a probability of 0.998. With its band
doubled it stays above 0.9 from 19:30, and with it tripled, from 20:23. The section-level
projection under the 2018 calibration passed 0.9 at 19:56 and 0.99 at 20:18. The markets passed
0.5 at 19:09 and 19:10, 0.9 at 20:24 and 0.99 at 21:00 and 21:02, with 87\% counted. Their timing was close to that
of a state-level extrapolation, which projects each state's remaining votes at its current
share and passed 0.9 at 20:36 and 0.99 at 20:49.

\begin{table}[t]
\centering
\caption{Runoff night: first minute from which P(Lula) stays above each threshold, and mean log
score of the outcome from 19:02 to 21:30 (Brasília time). $k$ multiplies the 2022-calibrated
standard deviation.}\label{tab:night}
\small
\begin{tabular}{lccc}
\toprule
 & 0.9 & 0.99 & Log score \\
\midrule
Live projection, own band & 19:15 & 19:15 & $-0.001$ \\
Live projection, band $\times 2$ & 19:30 & 20:34 & $-0.03$ \\
Live projection, band $\times 3$ & 20:23 & 20:55 & $-0.09$ \\
Section-level projection, $k=1$ & 17:08 & 19:06 & $-0.001$ \\
Section-level projection, $k=3$ & 19:06--19:25 & 20:05--20:13 & $-0.02$ to $-0.04$ \\
Section-level projection, $k=5$ & 19:36--20:01 & 20:29--20:50$^a$ & $-0.06$ to $-0.11$ \\
Section-level projection, 2018 calibration & 19:56 & 20:18 & $-0.11$ \\
State-level extrapolation & 20:36 & 20:49 & $-0.16$ \\
Headline extrapolation & 20:43 & 20:51 & $-0.32$ \\
Kalshi & 20:24 & 21:00 & $-0.21$ \\
Polymarket & 20:24 & 21:02 & $-0.22$ \\
\bottomrule
\end{tabular}
\par\smallskip
\begin{minipage}{0.92\textwidth}\footnotesize
The live projection starts at 19:15, so its log scores run from 19:15; over the same window the
markets score $-0.16$ (Kalshi) and $-0.17$ (Polymarket). Ranges for $k$ are over normal, $t_5$
and $t_3$ errors and minimum state errors of 0.0005 to 0.002; the 2018 row uses normal errors and
the 0.0005 minimum. The headline extrapolation projects the remaining votes at the headline
share. $^a$ Not reached before 21:30 with the 0.002 minimum. Sections public: 1.6\% at 19:15,
46\% at 20:24, 87\% at 21:00.
\end{minipage}
\end{table}

\subsection{The implied discount}

If the market used a projection's margin $M_t$ but a larger standard deviation $\sigma_t$, its
price would be $p_t=\Phi(M_t/\sigma_t)$, so $\sigma_t=M_t/\Phi^{-1}(p_t)$. When the market favours
the candidate the projection has behind, no positive $\sigma_t$ reconciles the two; we set
$\sigma_t=\infty$ in those minutes and keep them in the medians. We use minutes with
$|p_t-0.5|>0.05$, $0.01<p_t<0.99$ and $|M_t|>0.5$ points. Against the live projection, the
median ratio of $\sigma_t$ to the projection's standard deviation is 3.7 on Kalshi and 3.9 on
Polymarket over its snapshots. Against the section-level projection from 19:02, it is 2.1 on
both venues under the 2018 calibration and 8.5 (Kalshi) and 8.6 (Polymarket) under the 2022
calibration; calibrating on 2026's own errors, which no trader knew in advance, gives 7.2. A
single price cannot separate a larger standard deviation from a different expected margin; the
ratio measures the two together.

\subsection{Count updates and the headline}

From 19:47 to 20:24, between 14\% and 46\% counted, the headline showed Flávio ahead by up to one
point (Figure~\ref{fig:runoff}, bottom). The markets rose from 0.65 to 0.90 over that time, so
they did not follow the headline. Nor did they respond to the national files as they appeared:
regressing the change in the market's log-odds after each of the 198 files published from 19:00
to 21:10 on the file's change in projected margin and in headline share, no coefficient is
positive at any horizon from 10 seconds to 10 minutes (Appendix~\ref{app:data}). The state and
municipal files, which update first, may already have moved prices.

\subsection{Trading the gap}

Between 19:06 and 20:24, takers bought 2.19 million Lula contracts on Kalshi at an average price
of 0.78, and 10.8 million Lula-equivalent shares on Polymarket at 0.77. Lula won, so after
Kalshi's fee of about one cent the realized profit was 20.7 cents per contract. For the expected
profit we value each purchase at a benchmark's probability at the minute it was made. From
19:15, at the live projection's probability, it was 18.0 cents per Kalshi contract after fees,
12.5 cents with the band doubled and 3.8 cents with it tripled; on Polymarket, 21.2, 15.5 and 6.7
cents per share. From 19:06, under the 2018-calibrated section-level projection, it was 7.9 cents
on Kalshi and 10.0 cents on Polymarket. Under the state-level extrapolation it was 0.3 cents on
Kalshi, and under the headline extrapolation $-18.5$ cents. In 15 order-book snapshots in the window, the Polymarket
Lula book held a median of \$9,300 within two cents of the best ask (minimum \$1,100) and
\$27,500 within five cents (minimum \$6,600), at a median spread of one cent. Takers bought 10.8
million shares through it, so the book refilled as it was taken. We do not observe Kalshi's book
depth.

\section{Other Brazilian nights}\label{sec:contrasts}

\paragraph{First round, 2026.} The contracts paid on who finished first, so P(Lula) here is the
probability that Lula finished first. At 17:00 the markets gave Lula 0.68 to 0.69, following the
polls. Flávio led the headline from its first file at 17:23 and finished first by 2.0 points.
The section-level projection here uses the 2022 first round as reference, and the only earlier
pair on which to calibrate it, 2018 to 2022, had a different left candidate and an uneven swing.
Calibrated there, it passed 0.9 for Flávio only at 19:30, with 82\% counted. The markets passed
0.9 at 18:28, with 28\% counted, when that projection gave Flávio 0.62, and their log score from
17:25 to 21:30 was better ($-0.11$ against $-0.23$). A regional extrapolation of the count,
built like the runoff's state-level one, passed 0.9 at 18:40.

Table~\ref{tab:stage} compares the two 2026 nights at equal stages of the count. It measures
each projection's uncertainty by its errors on its own night, which no trader knew in advance.
In the first 5\% of the count the markets favoured the other candidate in most minutes on both
nights. From 5\% on, the markets' implied standard deviation was 6.7 to 8.3 times the
projection's in the runoff and 0.6 to 2.1 times in the first round.

\begin{table}[t]
\centering
\caption{Standard deviation of the final margin (points): market-implied (Kalshi) and the
section-level projection's, calibrated on each night's own errors (known only afterwards), by
share of sections public.}\label{tab:stage}
\small
\begin{tabular}{lcccccc}
\toprule
 & \multicolumn{3}{c}{Runoff, 25 Oct} & \multicolumn{3}{c}{First round, 4 Oct} \\
\cmidrule(lr){2-4}\cmidrule(lr){5-7}
\% public & Market & Projection & Ratio & Market & Projection & Ratio \\
\midrule
0--5 & \multicolumn{3}{c}{wrong side in 85\% of minutes} & \multicolumn{3}{c}{wrong side in 78\% of minutes} \\
5--20 & 7.38 & 0.89 & 8.3 & 1.28 & 2.20 & 0.6 \\
20--50 & 3.27 & 0.49 & 6.7 & 1.22 & 1.26 & 1.0 \\
50--90 & 2.09 & 0.27 & 7.8 & 1.05 & 0.49 & 2.1 \\
\bottomrule
\end{tabular}
\par\smallskip
\begin{minipage}{0.9\textwidth}\footnotesize
Medians over minutes with $|p_t-0.5|>0.05$, $0.01<p_t<0.99$ and $|M_t|>0.5$. ``Wrong side'':
the market favoured the candidate the projection had behind. Runoff 0--5 is mostly the
suspension.
\end{minipage}
\end{table}

\paragraph{Runoff, 2022.} Lula won by 1.80 points. Only Polymarket's automated market maker
traded, about \$70,000 that day, observed on a ten-minute grid. The market was at 0.89 in the
17:30 interval, on about \$1,000 of trades, and stayed above 0.9 from the 18:30 interval. Against
the section-level projection calibrated on its own night, its implied ratio is 5.8 (90\%
interval 4.1 to 8.1).

On count nights in Romania and Peru, against models calibrated on each night's own errors, the
same venues' ratios were close to one (Appendix~\ref{app:comparison}).

\section{Explanations}\label{sec:why}

We cannot test the following explanations.

\emph{A coarser model.} The markets' timing and accuracy were close to the state-level
extrapolation's (Table~\ref{tab:night}), and valued at it their buyers expected no profit. That
model ignores the geography of the count within states and is slow when early sections are
unrepresentative, as on runoff night. On the first-round night the headline was representative
from the start and a coarse model was enough.

\emph{Weight on 2018.} Under the 2018 calibration the markets were within about 0.1 of the
projection from 19:10 to 19:30, then fell 0.1 to 0.2 behind it until 20:20. This calibration
does not explain the suspension, when it too favoured Lula.

\emph{Distance and access.} The live projection and the volunteer tally appeared on a
Portuguese-language site with a small audience, and both venues were blocked in Brazil.

\emph{Costs and risks of trading.} A trader acting on the projection locked capital until the
result was declared and faced position limits on Kalshi. A contract paying on the declared
winner carries some risk of a contested result; a 5\% allowance for it lowers the
2022-calibrated ratio by about 0.7. Fees and the cost of capital also limit how close prices
get to 1, which matters late in the night, when prices exceeded 0.95, but not at 19:30, when the
markets were at 0.70 to 0.73 and the live projection at 0.998.

\section{Conclusion}\label{sec:conclusion}

On the night of Brazil's 2026 runoff, a projection published live gave the winner 0.998 more than
an hour before the markets reached 0.9. The markets ended on the right side, but their prices
imply about four times the live projection's uncertainty and about twice that of a projection
calibrated on the earlier runoff with the largest errors. Buying the winner at those prices had
a positive expected return under both projections, and none under a state-level extrapolation,
whose timing the markets matched. During the suspension the markets favoured the loser, though
the evidence against them was not available live in usable form. On the first-round night, when
the headline was representative from the start, they were faster than our projection.

\appendix
\section{Count nights in other countries}\label{app:comparison}

The same venues listed other count nights. None has Brazil's section-level timing, and the models
below are calibrated on each night's own errors. Table~\ref{tab:other} gives the implied ratios.

\paragraph{Romania, 18 May 2025.} Nicușor Dan beat George Simion by 7.2 points. Simion led the
headline from 2\% to 42\% of precincts, and Dan took the lead at 18:42 UTC. A same-precinct
swing on the first-round vote passed 0.9 for Dan between 18:17 and 18:34 UTC, and a model of
transfer rates from each first-round candidate between 18:38 and 18:58. The markets held Dan at
0.74 to 0.92 while Simion led the headline and passed 0.9 for good at 18:41 and 18:42. Treating
precincts as public when entered, the same-precinct model's log score from 18:17 was $-0.005$,
against $-0.020$ for the markets.

\paragraph{Peru, 7 June 2026.} The markets fell to 0.26 for Keiko Fujimori early on 8 June, with
46\% counted, while she led the headline by about five points and a department-level projection
gave her 0.74. They rose above 0.89 a day later and stayed there while her opponent, Roberto Sánchez,
led the headline, until Fujimori retook it on 11 June. She won with 50.1\% of the valid vote. Over the first 22 hours the implied ratio
against the department-level projection was 0.94.

\paragraph{United States, 5 November 2024.} Against the New York Times live model, shifted by its
difference from the markets before any votes were counted, the markets were more confident in six
of the seven swing states and less confident nationally and in Georgia. This comparison adjusts
for a prior, so it is not in the table.

\begin{table}[h]
\centering
\caption{Market-implied over model standard deviation, calibrated on each night's own errors.}\label{tab:other}
\small
\begin{tabular}{llc}
\toprule
 & Model & Ratio \\
\midrule
Brazil 2026 runoff & same sections & 7.2 \\
Brazil 2022 runoff & same sections & 5.8 \\
Romania 2025 runoff & same precincts; transfer rates & 1.0--1.8; 0.8--1.4 \\
Peru 2026 runoff, first 22 hours & departments & 0.94 \\
\bottomrule
\end{tabular}
\par\smallskip
\begin{minipage}{0.8\textwidth}\footnotesize
Medians over minutes; ranges over specifications and venues.
\end{minipage}
\end{table}

\section{Data checks}\label{app:data}

The code and the results behind every number in the paper are in this article's ancillary
files on arXiv (\texttt{anc/}).

\paragraph{Count path.} The archived national files (331 versions in the first round, 326 in the
runoff) never decrease in sections or in either candidate's votes. Their final versions equal the
section files' totals in both rounds (Lula 53,879,538 and Flávio 56,104,503 in the first;
64,256,232 and 59,826,041 in the runoff). Receipt and file generation are stamped by separate
systems, so in a check of 300 runoff versions a file with $n$ sections appears between 53 seconds
before and 32 seconds after the recorded receipt of the $n$-th section for the middle half of
versions (median 22 seconds after). In the runoff, versions
generated between 17:06:50 and 17:10 were served only at 19:00:50, and we treat them as public
from then.

\paragraph{Trades.} On Kalshi, summed trade sizes equal each market's reported volume in all 79
markets and each one-minute candle's volume in all 24,620 minutes with trades. Polymarket fills
from the platform's interface match the Polygon chain one for one on both election nights,
113,403 of 113,404 fills, the exception being a fill of 0.009 shares.

\paragraph{Count updates.} For each of the 198 national files published from 19:00 to 21:10 on
runoff night, we regress the change in the market's log-odds from the minute before the file to
$h$ seconds after it on the file's change in projected margin and in headline share, jointly,
with errors clustered by ten-minute block. No coefficient is positive at any horizon from 10
seconds to 10 minutes. Several are negative: at one minute on Kalshi, $-0.34$ per point of
projected margin (standard error 0.11) and $-0.03$ per point of headline share (0.01); at 30
seconds on Polymarket, $-0.33$ (0.16) and $-0.04$ (0.02).

\paragraph{Clocks.} Printing times are local and are converted to Brasília time by state. The
2022 market is on a ten-minute grid labelled by the start of each interval, so its times may be
up to nine minutes early.

\end{document}